\documentclass[aps,prl,groupedaddress,twocolumn]{revtex4-1}

\usepackage{float} 
\usepackage{bm} 
\usepackage{lipsum}
\usepackage[latin1]{inputenc}
\usepackage[T1]{fontenc}
\usepackage[english]{babel}
\usepackage{amsmath}
\usepackage{commath}
\usepackage{amssymb,amsfonts,textcomp}
\usepackage{color}
\usepackage{array}
\usepackage{hhline}
\usepackage{hyperref}
\usepackage{soul}
\hypersetup{pdftex, colorlinks=true, linkcolor=blue, citecolor=blue, filecolor=blue, urlcolor=blue, pdftitle=, pdfauthor=Benjamin Canals, pdfsubject=, pdfkeywords=}
\usepackage[pdftex]{graphicx}
\usepackage{booktabs}

\makeatletter
\newcommand\ps@Standard{
  \renewcommand\@oddhead{}
  \renewcommand\@evenhead{}
  \renewcommand\@oddfoot{}
  \renewcommand\@evenfoot{}
  \renewcommand\thepage{\arabic{page}}
}
\makeatother

\begin{document}

\title{Fragmentation of magnetism in artificial kagome dipolar spin ice}

\author{Benjamin Canals$^{1,2}$, Ioan Augustin Chioar$^{1,2}$, Van-Dai Nguyen$^{1,2}$, 
		Michel Hehn$^3$, Daniel Lacour$^3$, Fran\c cois Montaigne$^3$, 
		Andrea Locatelli$^4$, Tevfik Onur Mente\c s$^4$, Benito Santos Burgos$^4$ 
		\& Nicolas Rougemaille$^{1,2}$}
		



\begin{abstract}

Geometrical frustration in magnetic materials often gives rise to exotic, low-temperature states of matter, like the ones observed in spin ices. 
Here we report the imaging of the magnetic states of a thermally-active artificial magnetic ice that reveal the fingerprints of a spin fragmentation process.  
This fragmentation corresponds to a splitting of the magnetic degree of freedom into two channels and is evidenced in both real and reciprocal space. 
Furthermore, the internal organization of both channels is interpreted within the framework of a hybrid spin-charge model that directly emerges from the parent spin model of the 
kagome dipolar spin ice. 
Our experimental and theoretical results provide  insights into the physics of frustrated magnets and deepen our understanding of emergent fields through the use of tailor-made magnetism.
%

\end{abstract}

\maketitle

Frustration refers to the inability of a complex system to satisfy all its constraints simultaneously \cite{tou1979}. 
Geometrical frustration arises when these constraints are driven by the topology of an underlying lattice \cite{sad1999}. 
Examples of such systems can be found in a wide class of magnetic materials \cite{ram1994} and artificial magnetic architectures \cite{wan2006,qi2008,nis2013,hey2013} 
known as geometrically frustrated magnets. 
In some cases, these systems are characterized by highly fluctuating ground state manifolds, often referred to as spin liquids, and can exhibit unusual 
magnetic excitations such as fractional quasi-particles \cite{cas2008}.

The frustrated kagome Ising antiferromagnet is one of the first studied classical spin liquids, i.e. a strongly correlated magnetic model exhibiting only short-range spin-spin correlations down to the lowest temperatures \cite{syo51, kano53}. This short-range antiferromagnetic model can be mapped, one to one, onto a short-range ferromagnetic model, where geometrical frustration is provided by multiaxial Ising-like anisotropies \cite{moe1998}. The thermodynamics of both models is described by two temperature regimes: a high-temperature paramagnet and a low-temperature cooperative paramagnet, in which every triangular unit cell of the kagome lattice obeys the so-called (kagome) ice rule, meaning that two spins are pointing in or out of each triangle of the kagome lattice. This ice-like constraint on a triangle is akin to the original vertex water-ice rule \cite{ber1933} (hence their name, spin ices) which lies at the heart of the low-energy properties of square \cite{lie1967} and pyrochlore \cite{bra2001} spin ices.

Adding long-range dipolar interactions to the short-range frustrated multiaxial ferromagnet drastically modifies its low-temperature behavior \cite{mol2009}. The first spin ice manifold (SI1) survives over a finite temperature range and is then followed, at lower temperatures, by a second spin ice manifold (SI2), in which spin loop fluctuations coexist with an effective magnetic charge crystal. Eventually, a N\'eel-like ordering (Long Range Order - LRO) occurs at the lowest temperatures. 
In both the SI1 and SI2 phases, the kagome ice-rule is obeyed. However, the magnetic charges, associated to the fractionalization of each spin into opposite pole pairs, are in a paramagnetic state in the SI1 phase, while they crystallize in the SI2 phase \cite{mol2009,che2011}. Although more constrained, the SI2 phase remains macroscopically degenerated, as shown by its finite entropy.

The puzzling aspect of the SI2 phase is the emergence of the magnetic charge crystallization, which is a priori not encoded in the underlying dipolar spin ice model \cite{mol2009,che2011}. Experimental evidences of charge crystallites in artificial kagome spin ices have been also reported \cite{rou2011,chioar1,mon2014,drisko2015,zha2013,ang2015} and interpreted with the use of a phenomenological spin-charge model \cite{zha2013}. 
Here, we provide a theoretical framework that allows us to reveal the microscopic origin of this emerging charge organization.
Besides, in the particular case of the kagome dipolar spin ice, the formation of an antiferromagnetic charge crystal is a direct consequence of the recent proposal 
of spin fragmentation \cite{bro2014}. 
Furthermore, by using thermally active kagome arrays of nanomagnets \cite{chioar1,mon2014,drisko2015,zha2013,ang2015,far2013,far2014}, we evidence experimental signatures of this fragmentation of magnetism.

\section*{Results}
\textbf{Ab initio coding of emergence.}
In this section, we discuss how the magnetic charge at the vertices is encoded into the Hamiltonian of the dipolar kagome spin ice model and we reveal why these charges 
crystallize at low temperature. To do so, we consider the dipolar Hamiltonian
\begin{equation}
H_{dip} = \frac{D}{2} \cdot \sum_{(i,j)} [ \frac{\mathbf{S_i} \cdot \mathbf{S_j}}{r_{ij}^{3}} - \frac{3 \cdot (\mathbf{S_i} \cdot \mathbf{r}_{ij}) \cdot (\mathbf{S_j} \cdot \mathbf{r}_{ij})}{r_{ij}^{5}}]
\label{eq:original_ham}	
\end{equation} 
\noindent
where $D$ is the dipolar constant, $r_{ij}$ the distance between sites $i$ and $j$, and $\mathbf{S}_i = \sigma_i \mathbf{e}_i$ an Ising-like spin residing on site $i$ and pointing along its local anisotropy axis (see Figure \ref{fig:thermodynamics}a). Once this model is brought to low temperatures, ice rules are unanimously fulfilled or, 
equivalently, the total magnetic charge at each vertex of the honeycomb lattice is $\pm 1$ (see Figure \ref{fig:thermodynamics}b). Since the vertex charges are defined as local sum of the individual magnetic poles of the fractionalized spins, it is this spin/charge correspondence, together with the particular structure of the dipolar interactions in this lattice, which allow the rewriting of the microscopic spin model. The Hamiltonian then becomes a hybrid spin-charge model with only Ising-like variables, 
\begin{equation}
\label{eq:rewritten_ham}
H_{dip} = - \tilde{J} \sum_{\langle i,j \rangle} \mathbf{S}_i . \mathbf{S}_j
						- \tilde{\kappa} \sum_{\langle u,v \rangle} Q_u . Q_v
						+ O \left( 1 / r^3 \right)_{r \ge 2 r_{\textrm{nn}}}
\end{equation}
\noindent
where the first term is a short-range spin ice kagome ferromagnet (i.e. an Ising spin liquid), while the second term corresponds to an antiferromagnet of magnetic charges $Q$ on an hexagonal lattice (see Methods section). The remaining terms correspond to the longer-range part of the dipolar spin model. The sum of these last terms is, algebraically speaking, absolutely convergent and they do not interfere with the physics driven by the first two contributions, except at very low temperatures where they lead to the N\'eel ordering (LRO).

This exact mapping of eq. \ref{eq:original_ham} onto eq. \ref{eq:rewritten_ham} explains why charge crystallization is observed in the SI2 phase. As $\left| \tilde{J} \right| > \left| \tilde{\kappa} \right|$, the first term ensures that kagome ice rules are fulfilled when the temperature is lowered, and the model fluctuates within the first spin ice manifold, SI1. Because the Ising charge-charge interaction is unfrustrated and is fully compatible with the SI1 constraint, it leads to charge crystallization when the temperature is further reduced, i.e. it selects a sub-manifold of the SI1 set, the SI2 manifold. We emphasize that this exact algebraic mapping of the original spin model onto a hybrid spin-charge Hamiltonian is one of the very few examples of ab initio coding of emergence: the collective organization of magnetic charges is directly related to the original microscopic spin degree of freedom.

\noindent
\textbf{Fragmentation of magnetism in the SI2 phase.}
Once this model is brought into the SI2 phase, magnetic charges at each vertex are constrained to $Q = \pm 1$. As noted by Brooks-Bartlett and collaborators \cite{bro2014}, a formal analogy with electrostatics can be made and this distribution of unitary magnetic charges can be related to the divergence of the local magnetization, which is proportional to the sum of the spins pointing inwards and outwards of the corresponding triangle of the kagome lattice. This means that
\begin{equation}
\mathbf{\nabla} . \left( \sum_{i \in v} \mathbf{S_i}  \right) \propto Q_{v}
\end{equation}
\noindent
where $\mathbf{S_i}$ is one of the 3 spins participating to the vertex $v$ and $Q_{v}$ the associated unitary charge. Applying a lattice Helmholtz-Hodge decomposition on the vector field $\{\sum_{i \in v} \mathbf{S_i} \}_{v}$ results in a splitting into a curl-free and a divergence-free contribution, 
\begin{equation}
\sum_{i \in v} \mathbf{S_i} = \mathbf{\nabla}( \phi_v ) + \mathbf{\nabla} \times \bm{\mathcal{C}_v}
\end{equation}
\noindent
where $\phi$ is a scalar field and $\bm{\mathcal{C}}$ a vector field. From this decomposition, it is clear that only the scalar field $\phi$ carries information on the magnetic charge as 
\begin{equation}
\mathbf{\nabla} . \left( \sum_{i \in v} \mathbf{S_i} \right) = \Delta \phi_v \propto Q_{v}
\end{equation}
\noindent
It is worth noting that such decomposition is simply an algebraic rewriting and can be applied to any vector field defined on a lattice.

The remarkable property of this model is the decoupling at low temperatures of the divergence-free and divergence-full channels. Within the SI2 region, the scalar field $\phi$ orders because of the $Q_u . Q_v$ term in the Hamiltonian, which leads to the charge crystallization. On the other hand, the first two terms of the hybrid spin-charge Hamiltonian do not impact the vector field $\bm{\mathcal{C}}$, leaving it free to fluctuate. This last property, associated to its natural divergence free nature, defines a so-called Coulomb phase \cite{hen2010}. This exotic state of matter, within which order and disorder coexist, is very unusual as both aspects of magnetic organization are carried by the same degree of freedom, hence the name fragmentation. 

\noindent
\textbf{Looking for signatures of spin fragmentation.}
This theoretical framework provides insights into how fingerprints of this fragmentation process can potentially be revealed in experiments. While Figure \ref{fig:helmholtz_hodge} details the fragmentation process in real space, a more straightforward way to visualize it is by plotting the magnetic structure factor $S(q)$, i.e. the Fourier transform of spin-spin correlations. In reciprocal space, the static divergence-full channel is revealed through magnetic Bragg peaks, while the divergence-free channel appears as a structured diffuse signal \cite{bro2014}. It is the coexistence of both types of signal which demonstrates spin fragmentation. 
The temperature dependence of the 2-dimensional (2D) magnetic structure factor for the dipolar spin ice model has been computed using Monte Carlo simulations (see Methods section) 
and is reported in Figure \ref{fig:thermodynamics}c along with the temperature dependence of the entropy and  specific heat. 
In the following analysis of our measurements, it is this coexistence of Bragg peaks and structured diffuse background signal that we use to evidence the spin fragmentation process experimentally. Note that charge crystallization corresponds to a rather counter-intuitive phenomenon. The magnetic charge crystal is the emerging description of an antiferromagnetic all-in / all-out fragmented spin ordering, as depicted in Figure \ref{fig:helmholtz_hodge}, in a nevertheless ferromagnetic spin model. Real space imaging of artificial spin ice systems allows direct visualization of this counter-intuitive phenomenon.

\noindent
\textbf{Experimental evidence of spin fragmentation.}
Evidencing the spin fragmentation process in artificial kagome spin ice is challenging, mainly because of the experimental difficulty to bring such systems into their low-energy manifolds, where collective phenomena emerge. In the following, we show however that this can be done in thermally-active, kagome arrays of connected Gd$_{0.3}$Co$_{0.7}$ nanomagnets (see Methods section, Supplementary Note 1 and Supplementary Figure 1), 
which have been studied using XMCD-PEEM magnetic imaging (see Methods section). A typical magnetic image is reported in Figure \ref{fig:xpeem}a. Black and white contrasts allow the determination of the magnetization direction within each single nanomagnet. Hence, the overall spin configuration of the array is deduced together with the distribution of the associated magnetic charges at the vertex sites.

Using the spin configuration of the whole lattice obtained from our XMCD-PEEM measurements, we can compute the magnetic structure factor and compare it to the one predicted by Monte Carlo simulations at a similar effective temperature (the effective temperature of our array is estimated by comparing the measured spin-spin correlation coefficients to their thermodynamic values \cite{chioar1,chioar2,rou2011}, see Methods section). The result is reported in Figure \ref{fig:exp_vs_theo}. 
At first sight, there is a fairly good qualitative agreement between the experimental [Figure \ref{fig:exp_vs_theo}a, top] and theoretical [Figure \ref{fig:exp_vs_theo}a, bottom] 2D maps of the magnetic structure factor. 
The most striking feature of the experimental map consists in the clear fingerprints of a spin fragmentation process which are twofold:  appearance of Bragg peaks (see black circles) 
and a structured diffuse background signal accounting for the disordered phase of the divergence-free component (see yellow regions in the 2D maps).
We emphasize again that it is this coexistence which demonstrates spin fragmentation. The sole presence of charge ordering is not sufficient to evidence spin fragmentation, as the key aspect of this phenomenon is the emergent decoupling of the divergence-free and divergence-full spin channels. For example, the LRO ground state or a saturated magnetic configuration would directly relate charge crystallization to spin ordering (i.e. there is no fragmentation of magnetism through the decoupling of the two channels).

This qualitative agreement between the experimental and theoretical maps can be made more quantitative.
Figure \ref{fig:exp_vs_theo}b provides a comparison between the experimental and the theoretical magnetic structure factors along a q-scan through the reciprocal space, passing through the fragmentation peak [black circle in Fig. \ref{fig:exp_vs_theo}a]. Because the experimental statistics is low, we have reported the theoretical Monte Carlo signal along the q-scan with its standard deviations in order to quantify the likelihood of the dipolar spin ice model (DSI) to describe our measurements. It appears that the DSI model captures quantitatively the fragmentation process. The experimental image also displays several regions of alternating $+1$ and $-1$ magnetic charges, pointing to incipient charge ordered crystallites. This sample has therefore not reached the thermodynamic SI2 phase, within which a unique crystallite would be expected, and is consistent with the effective temperature $T/J_{nn}=0.051$ deduced for the spin-spin correlator analysis, that place the array deep into the SI1 phase. Thanks to the real space imaging of this artificial magnet, the lattice Helmholtz-Hodge decomposition can be performed for each crystallite. One of these crystallites is highlighted by an orange rectangle and better illustrated in Figure \ref{fig:xpeem}b. Along with the black arrows that indicate the local spin directions, we use again the red/blue color code for each kagome triangle to represent the vertex charge state. In this selected region, there is no spin order, although the magnetic charge has crystallized \cite{zha2013, mon2014,chioar1,drisko2015} and is embodied in a fragmented all-in / all-out antiferromagnetic spin ordering [see Fig. \ref{fig:xpeem}c]. The array has thus been brought, locally, into the SI2 phase.

Finally, we emphasize that our observation of a spin fragmentation process is not limited to the case of our thermally active GdCo alloy spin ice system. Very similar results have been obtained with more conventional, athermal, permalloy-based kagome arrays, that we demagnetized using a field protocol prior to their imaging in a Magnetic Force Microscope. In these types of samples, we also observed the coexistence of Bragg peaks and a structured diffuse background signal in the 2D maps of the magnetic structure factor (see Supplementary Note 2, Supplementary Table 1 and Supplementary Figure 2), thus proving the generality of the concept and the capability to measure it experimentally using different materials and demagnetization protocols.

\section*{Discussion}
In conclusion, we report the experimental signatures of a magnetic-moment fragmentation process in a thermally-active artificial magnet and provide a theoretical framework to interpret this emerging phenomenon. We emphasize that the long-range dipolar nature of the interactions in artificial magnetic ices is at the heart of both our experimental and theoretical results. Achieving a complete tuning of the cooling procedure now becomes an ultimate goal as it would provide a statistical physics laboratory, paving the way for engineered magnetic structures and their use to understand, realize and control new states of matter, be they artificial or not.

\bigskip

\section*{Methods}
\subsection{Sample fabrication.}
Our samples were fabricated on Si substrates from full films grown by ultra-high vacuum sputtering (base pressure in the $10^{-9}$ mbar range). The final stack has the following composition: Si//Ta (5 nm)/Gd$_{0.3}$Co$_{0.7}$ (10 nm)/ Ru (2.6 nm). The 5 nm-thick Ta layer is used as a buffer for the subsequent growth of the magnetic GdCo film, which is finally capped with a 2.6 nm-thick Ru layer. The capping material and its thickness have been optimized in order to protect the sample from oxidation and against chemical treatments during the lithography process, while still keeping high the magnetic contrast in imaging conditions. The Gd$_{0.3}$Co$_{0.7}$ material is a ferrimagnetic alloy characterized by a Curie temperature ($T_C$) of about 475 K, which was adjusted by co-sputtering Co and Gd in DC mode to control the alloy composition. Additional information on the magnetic properties of the GdCo alloy is provided in the Supplementary Note 1. The arrays that we fabricated using e-beam lithography and ion beam etching are composed of 342 nanomagnets having typical dimensions of $500 \times 100 \times 10 $ nm$^{3}$. At room temperature, each nanomagnet has a magnetization pointing along the long axis of the element, due to shape anisotropy, and can thus be considered as an Ising pseudo-spin.

\subsection{Thermal annealing protocol.}

Experimentally, the arrays are first saturated using an external magnetic field to set the initial spin configuration in a well-defined state. The arrays are then heated up above the Curie temperature of the material by passing a current through a W filament underneath the sample stage. The cooling procedure was performed as follow: when reaching the targeted temperature the filament current was quickly switched off to avoid the presence of an Oersted field while cooling down the sample through $T_C$. The typical cooling time is 30 minutes and is basically limited by the thermal dissipation into the sample holder. After cooling down the sample back to room temperature, the resulting spin configurations of the arrays are imaged using X-xay PhotoEmission Electron Microscopy (PEEM) combined with X-ray Circular Magnetic Dichroism (XMCD). Measurements were done at the Nanospectroscopy beamline of the Elettra synchrotron, Trieste, Italy. 

\subsection{Estimation of an effective temperature.}

Although in a frozen magnetic state (i.e. the magnetic configuration does not evolve anymore once the sample is at room temperature) when being imaged, an effective temperature can be associated to the spin configuration resulting from our thermal treatment. To do so, we proceed similarly to what we did in previous works by employing a standard deviation analysis through the use of a ''spread-out'' function based on the spin-spin correlators \cite{chioar1,chioar2}. 
This "spread-out" function is defined as:  $K(T/J_{\rm{nn}})=\sqrt(\sum_{j}^{} (C_{j}^{exp}-C_{j}^{MC}(T/J_{\rm{nn}}))^2)$ where $C_{j}^{exp}$ represent the experimental correlations, while $C_{j}^{MC}(T/J_{\rm{nn}})$ are the average Monte Carlo correlations at a given temperature $T/J_{\rm{nn}}$, with $j$ ranging from 1, the nearest-neighbor correlation, up to 7 (the 7th next nearest-neighbors).
For our set of experimental values, this function can be computed over the entire range of Monte Carlo temperatures.
The minimum of $K(T/J_{\rm{nn}})$ defines the effective temperature.
Additional works on artificial spin ice systems where an effective temperature is associated to an arrested (i.e. frozen) microstate can be found for example in Refs.\cite{mor2013,kap2014}.

\subsection{Monte Carlo simulations.}
The Monte Carlo simulations were performed on $12 \times12 \times 3$ kagome lattice sites with periodic boundary conditions (PBC). This geometry corresponds to the PBC cluster having the closest number of sites to the one of the finite experimental realization, and compatible with the periodicity of the Long Range Order expected at the lowest temperatures. $10^{4}$ modified Monte Carlo steps (MMCS) are used for thermalization and measurements are computed over $10^{4}$ MMCS, where one MMCS involves local spin-flips as well as global loop-flips, such that the analysis of the integrated correlation time calculated on the fly ensures stochastic decorrelation between measurements. We note that simulations performed for a network of 342 spins with free boundary conditions yield very similar results. Furthermore, the same ground-state configuration is found for this particular finite system size as in the infinite network case. Both types of simulations ensure that all the physics at stake, be it experimental or theoretical, does not depend on the cluster geometry or on the boundary conditions.

\subsection{Dipolar hamiltonian.}
Nanomagnets interact through the magnetostatic interaction. They possess a strong anisotropic shape (aspect ratio of about 5) and are well approximated by Ising-like variables oriented along their long axis. As shown in Ref.\cite{rou2011}, the point dipole approximation of the magnetostatic terms is valid only beyond nearest neighbors. For the closest elements, their shape and proximity prevent the dipolar approximation to be valid. It appears however that multipolar terms can actually be taken into account correctly, provided that the nearest neighbor dipolar term is enhanced by an extra coupling $J_1$. The full Hamiltonian describing this system then reads
\begin{equation}
H = - J_1 \sum_{\langle i,j \rangle} \mathbf{S}_i . \mathbf{S}_j 
			 + \frac{D}{2} \cdot \sum_{(i,j)} [ \frac{\mathbf{S_i} \cdot \mathbf{S_j}}{r_{ij}^{3}} - \frac{3 \cdot (\mathbf{S_i} \cdot \mathbf{r}_{ij}) \cdot (\mathbf{S_j} \cdot \mathbf{r}_{ij})}{r_{ij}^{5}}]
\end{equation}
\noindent
where $D$ is the dipolar constant, $r_{ij}$ the distance between sites $i$ and $j$, and $\mathbf{S}_i = \sigma_i \, \mathbf{e}_i$ the spin residing on site $i$, with $\sigma_i$ an Ising variable and $\mathbf{e}_i$ one of the three possible anisotropy directions of the kagome lattice. To scale the temperature, we rely on the effective coupling $J_{\textrm{nn}} = J_1/2 + 7D/4$. This coupling quantifies the nearest-neighbor effective interaction between two nanomagnets, i.e. the absolute temperature at which the model is expected to enter the first spin ice manifold, SI1. Following our previous work \cite{rou2011}, $J_1$ is chosen such that $J_{nn}=5D$ to account for the multipolar terms mentioned above.

\subsection{Ab initio coding of emergence.}
Given the fractionalization of the spins into opposite magnetic poles, the total magnetic charge for each kagome vertex can be written as the sum of three individual charge contributions [see Figure \ref{fig:Supp1}a]. Using the scalar values ($\sigma_i = \pm 1$) that define the orientation of each lattice spin ($\mathbf{S_i}$) with respect to their local anisotropy axis ($\mathbf{e_i}$), i.e. $\mathbf{S_i}= \sigma_i \, \mathbf{e_i}$, the value of each vertex charge can be expressed as the sum of these local spin scalars,
\begin{equation}
Q_{\Delta} = -\sum_{i\in\Delta} \sigma_i
\quad
\textrm{and}
\quad
Q_{\nabla} = \sum_{i\in\nabla} \sigma_i \; ,
\label{eq:eq1}
\end{equation}
where $Q_{\Delta}$ and $Q_{\nabla}$ are the values of the vertex charges of a ${\Delta}$-type/${\nabla}$-type kagome triangle, while the $-$ sign ensures global charge neutrality. A nearest-neighbor charge-charge term, $Q_u \cdot Q_v$, where $u$ and $v$ are the indices of the hexagonal lattice, involves the product of two $Q_{\Delta, \nabla}$ charges, which can be expanded into a summation of $\sigma_i . \sigma_j$ pairs. Furthermore, by conveniently arranging the resulting pairs into their corresponding correlation classes [see Figure \ref{fig:Supp1}b], this charge-charge term can be expressed as a linear combination of the first three pairwise scalar spin correlators, 
\begin{equation}
(Q_{u}Q_{v}) = - 1 + 8 \cdot C_{\alpha \beta} + 4 \cdot C_{\alpha \gamma} - 2 \cdot C_{\alpha \nu}  	 
\label{eq:eq4}		
\end{equation}
This relation between the nearest-neighbor charge correlator and the spin correlators is the cornerstone of the hybrid spin-charge model description, which can be tailored out of the dipolar Hamiltonian,
\begin{equation}
H_{dip} = \frac{D}{2} \cdot \sum_{(i,j)} [ \frac{\mathbf{S_i} \cdot \mathbf{S_j}}{r_{ij}^{3}} - \frac{3 \cdot (\mathbf{S_i} \cdot \mathbf{r}_{ij}) \cdot (\mathbf{S_j} \cdot \mathbf{r}_{ij})}{r_{ij}^{5}}]
\label{eq:eq5}	
\end{equation} 
\noindent
where $\mathbf{r_{ij}}$ stands for the relative position vector between the two spins $\mathbf{S_i}$ and $\mathbf{S_j}$, and $D$ is the dipolar coupling constant. Note that the dipolar Hamiltonian involves all spin-spin correlations, in particular the ones needed in eq. \ref{eq:eq4}. Expanding the dipolar couplings and grouping them appropriately allows to rewrite the Hamiltonian as:
\begin{widetext}
\begin{equation}
H_{dip} = - \tilde{J} \sum_{\langle i,j \rangle} \mathbf{S_i} \cdot \mathbf{S_j} - \tilde{\kappa} \sum_{\langle u,v \rangle} Q_u \cdot Q_v + 
                \frac{D}{2} \cdot \sum_{\substack{(i,j)\\r_{ij} \geq 2r_{nn}}} [ \frac{\mathbf{S_i} \cdot \mathbf{S_j}}{r_{ij}^{3}} - \frac{3 \cdot (\mathbf{S_i} \cdot \mathbf{r}_{ij}) \cdot (\mathbf{S_j} \cdot \mathbf{r}_{ij})}{r_{ij}^{5}}]
\end{equation}
\end{widetext}
\noindent
which will be written, for the sake of clarity,
\begin{equation}
H_{dip} = - \tilde{J} \sum_{\langle i,j \rangle} \mathbf{S_i} \cdot \mathbf{S_j} - \tilde{\kappa} \sum_{\langle u,v \rangle} Q_u \cdot Q_v  + 
O \left( 1 / r^3 \right)_{r \ge 2 r_{\textrm{nn}}}
\label{eq:eq19}		
\end{equation}
\noindent
with $\tilde{J} = 2 D_{\textrm{eff}} (7/4 + 5\sqrt{3}/9)$, $\tilde{\kappa} = -5 D_{\textrm{eff}} \sqrt{3}/18$, $D_{\textrm{eff}} = D/r_{\rm{nn}}^3$ and $r_{\rm{nn}}$ the nearest-neighbor distance. Since the coupling constants $\tilde{J}$ and $\tilde{\kappa}$ are positive and negative, respectively, they correspond to a ferromagnetic coupling between the nearest neighboring spins and to an antiferromagnetic coupling between the nearest neighboring vertex charges. As mentioned in the manuscript, the first one ensures that ice rules are fulfilled, while the second one leads to charge crystallization at lower temperatures, as $| \tilde{J} | > \left| \tilde{\kappa} \right|$. Note also that the charge-charge term is fully compatible with the spin-spin coupling and is not frustrated as it is defined on the hexagonal lattice. The remaining longer-range couplings play an important role as well as they ultimately lead to the magnetic ordering at the lowest temperatures.
This magnetic ordering does not interfere with the first two terms of the hybrid spin-charge model: 
it is compatible with the kagom\'e ice rules, the magnetic charge crystal, and the absolute convergence of the series of the longer-range couplings shows that this ordering 
takes place at temperatures lower than $\left| \tilde{\kappa} \right|$ (and hence than $| \tilde{J} |$ as well).

\section*{Acknowledgments}

This work was partially supported by the Region Lorraine and the Agence Nationale de la Recherche through Project No. ANR12-BS04-009 'Frustrated'. I.A.C. acknowledges financial support from the Laboratoire d'Excellence LANEF in Grenoble.
The authors also acknowledge support from the Nanofab team at the Institut NEEL and warmly thank S. Le-Denmat and O. Fruchart for technical help during AFM/MFM measurements.

\section*{Author contributions}

B.C. and N.R. conceived the projet. M.H. and D.L. were in charge of the thin films growth and optimization of magnetic properties. F.M. patterned the samples. F.M., D.L., M.H. and N.R. participated to the XMCD-PEEM experiments. The XMCD-PEEM measurements were carried out by N.R., A.L., O.T.M, B.S.B. Complementary measurements on athermal kagome spin ices were done by V.D.N. who was in charge of both sample preparation and imaging. I.A.C. and B.C. developed the Monte Carlo approach. B.C. and N.R. analyzed the data and wrote the manuscript.

\section*{Additional information}

Supplementary information accompanies this paper.

\section*{Competing financial interests}
The authors declare no competing financial interests.

\begin{widetext}

\begin{figure}
\centering
\includegraphics[width=13cm]{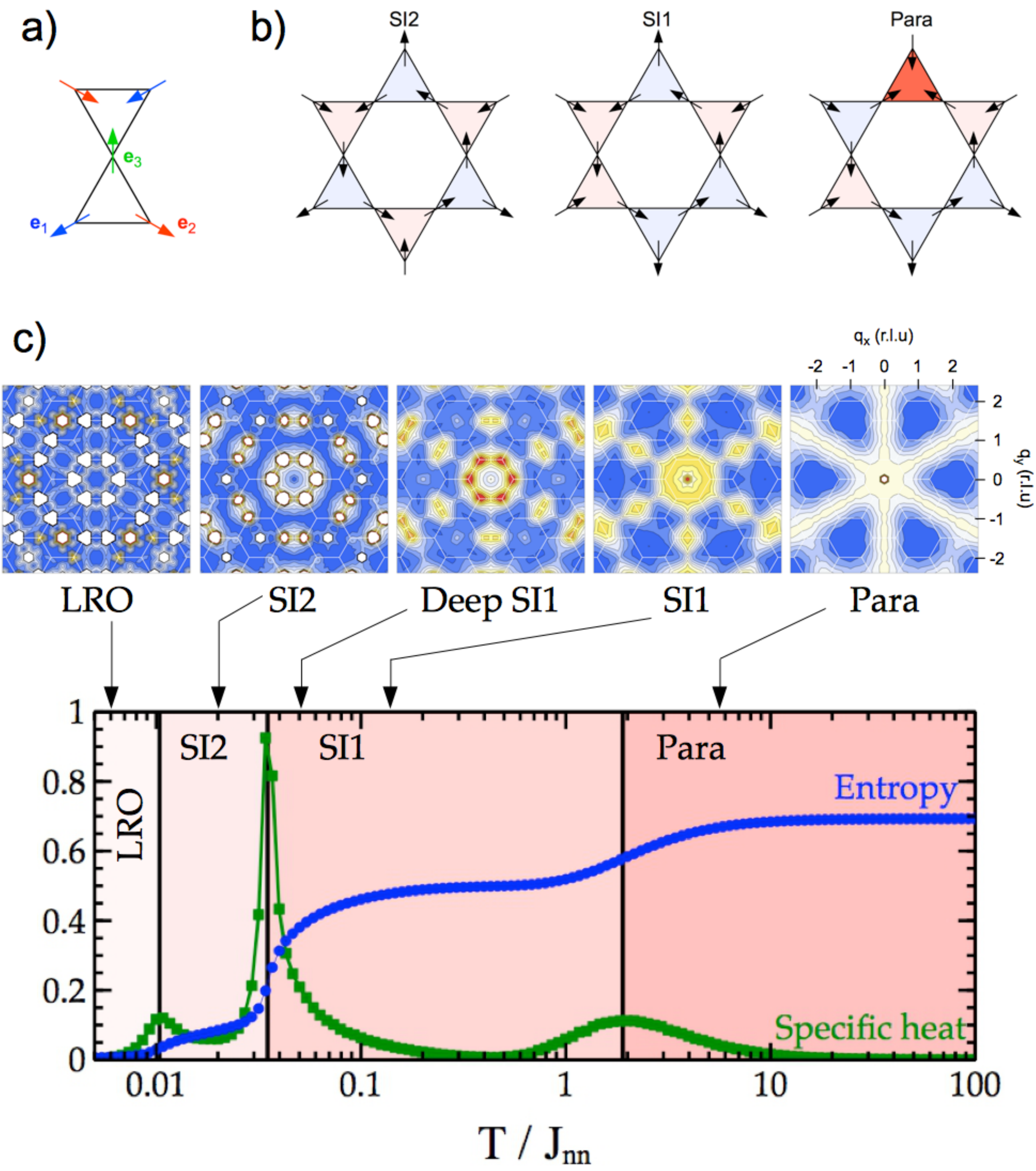} 

\caption{{\bf Thermodynamics of the kagome dipolar spin ice.} (a) The three possible anisotropy directions in the kagome lattice are denoted by the $\mathbf{e}_{i}$ unit vectors, which allow the description of the Ising-like spins as $\sigma_{i}$ scalar variables. (b) Three examples of magnetic configurations belonging to the spin ice 2 (SI2), spin ice 1 (SI1) and paramagnetic regimes. Each spin at the vertex of the kagome lattice can be seen as a magnetic dipole and fractionalized into an opposite pole pair. By conventionally taking the head of the arrow to be a +1 magnetic charge and the tail as a -1 magnetic charge, a total magnetic charge can be attributed to each triangle, i.e. to each vertex of the kagome lattice, by summing the 3 elementary charges of the fractionalized spins. A total $Q=+1/-1$ is depicted by a light red/blue triangle, respectively. The paramagnetic phase also features $\pm 3$ charges, as indicated for instance by a dark red triangle. As soon as the system enters its spin ice manifolds, ice rules are unanimously obeyed, which translate into the presence of only unitary charges. While these charges are disordered within the SI1 phase, they eventually crystallize antiferromagnetically in the SI2 phase, as depicted by the alternation of red-blue triangles.
(c) The simulated temperature dependencies of the entropy and the specific heat of the kagome dipolar spin ice model. Magnetic structure factor [$S(q)$] maps associated to each phase are provided and their corresponding normalized temperatures $T/J_{\textrm{nn}}$ are 5.815 (paramagnetic), 0.131 (SI1), 0.051 (deep SI1), 0.020 (SI2) and 0.006 (LRO), where $J_{\textrm{nn}}$ is the effective nearest-neighbour interaction.
\label{fig:thermodynamics}}

\end{figure}

\begin{figure}
\centering
\includegraphics[width=13cm]{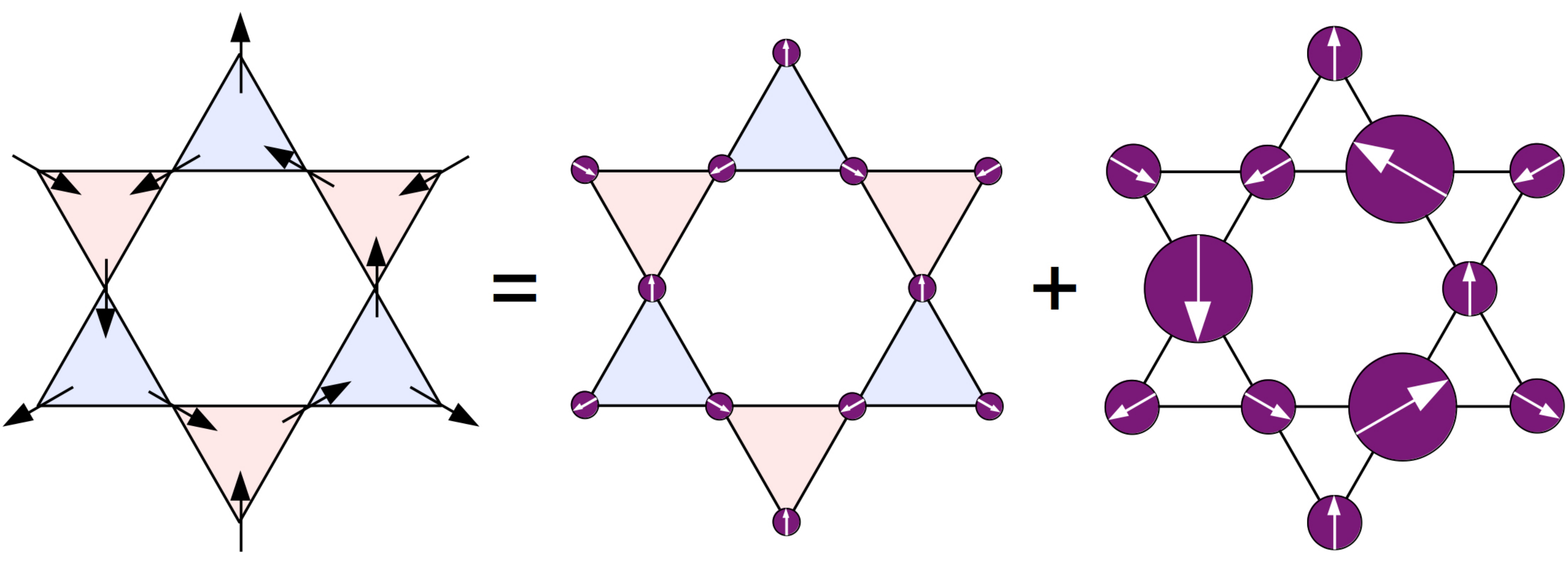} 

\caption{{\bf Illustration of the spin fragmentation process on a configuration belonging to the SI2 phase.} Spins are represented by black arrows, while the associated magnetic charges at the kagome vertices are represented by a red/blue triangle, corresponding to a $+1$/$-1$ charge state, respectively. By applying a Helmholtz-Hodge decomposition over the entire network, each spin of the lattice fragments into two channels, such that the magnetic configuration decomposes into a divergence-full and a divergence-free field. Note that, for the two fields, both the spin direction and the spin length change. To better visualize the fragmentation process, fragmented spins are represented by purple circles of diameters 1/3, 2/3, and 4/3, according to the moment of the fragmented part. This type of decomposition can be performed on {\it any} spin configuration belonging to the SI2 phase, for which the divergence-full channel is always the same and independent of the initial spin configuration, while the divergence-free channel can fluctuate, ensuring the magnetic equivalent of a Kirchhoff law.
\label{fig:helmholtz_hodge}}

\end{figure}

\begin{figure}
\centering
\includegraphics[width=13cm]{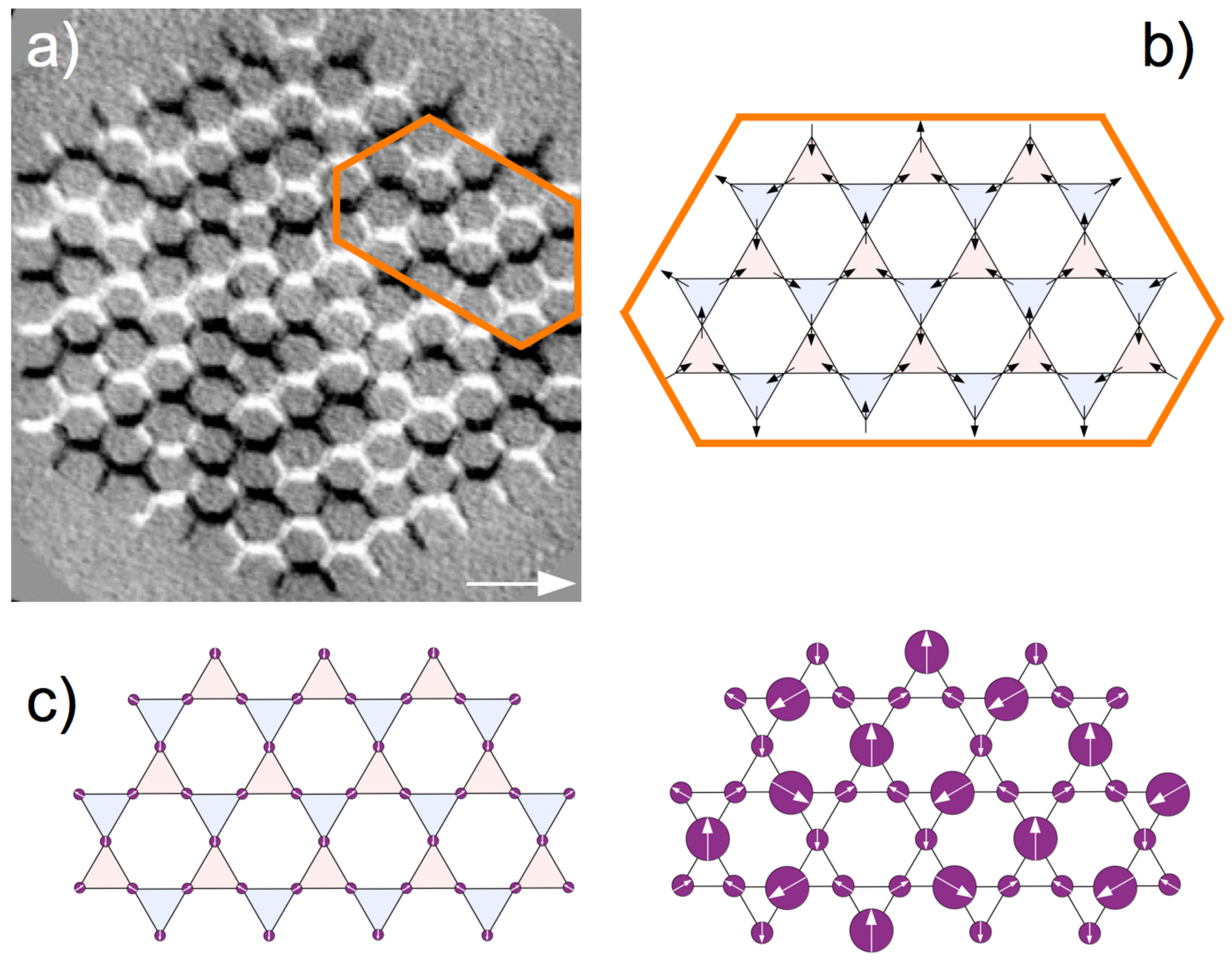}

\caption{{\bf Experimental evidence in real space of spin fragmentation in a thermally active kagome dipolar spin ice.} (a) 10x11 $\mu m^2$ XMCD-PEEM image of an artificial kagome array. 
The black and white contrasts give the local direction of the magnetization within each individual nanomagnet. 
The white arrow indicates the direction of the incident X-ray beam.
The orange hexagon highlights a region of the array where perfect charge ordering is observed. 
(b) The local spin configuration within the orange hexagon deduced from the XMCD-PEEM image. 
Spins are represented by black arrows, while the associated magnetic charges at the kagome vertices are represented by a red/blue triangle, 
corresponding to a +1/-1 charge state, respectively.
(c) Helmholtz-Hodge decomposition performed on the local spin configuration shown in the orange hexagon.
Fragmented spins are represented by purple circles of diameters 1/3, 2/3, and 4/3, according to the moment of the fragmented part.
\label{fig:xpeem}
}

\end{figure}

\begin{figure}
\centering
\includegraphics[width=14cm]{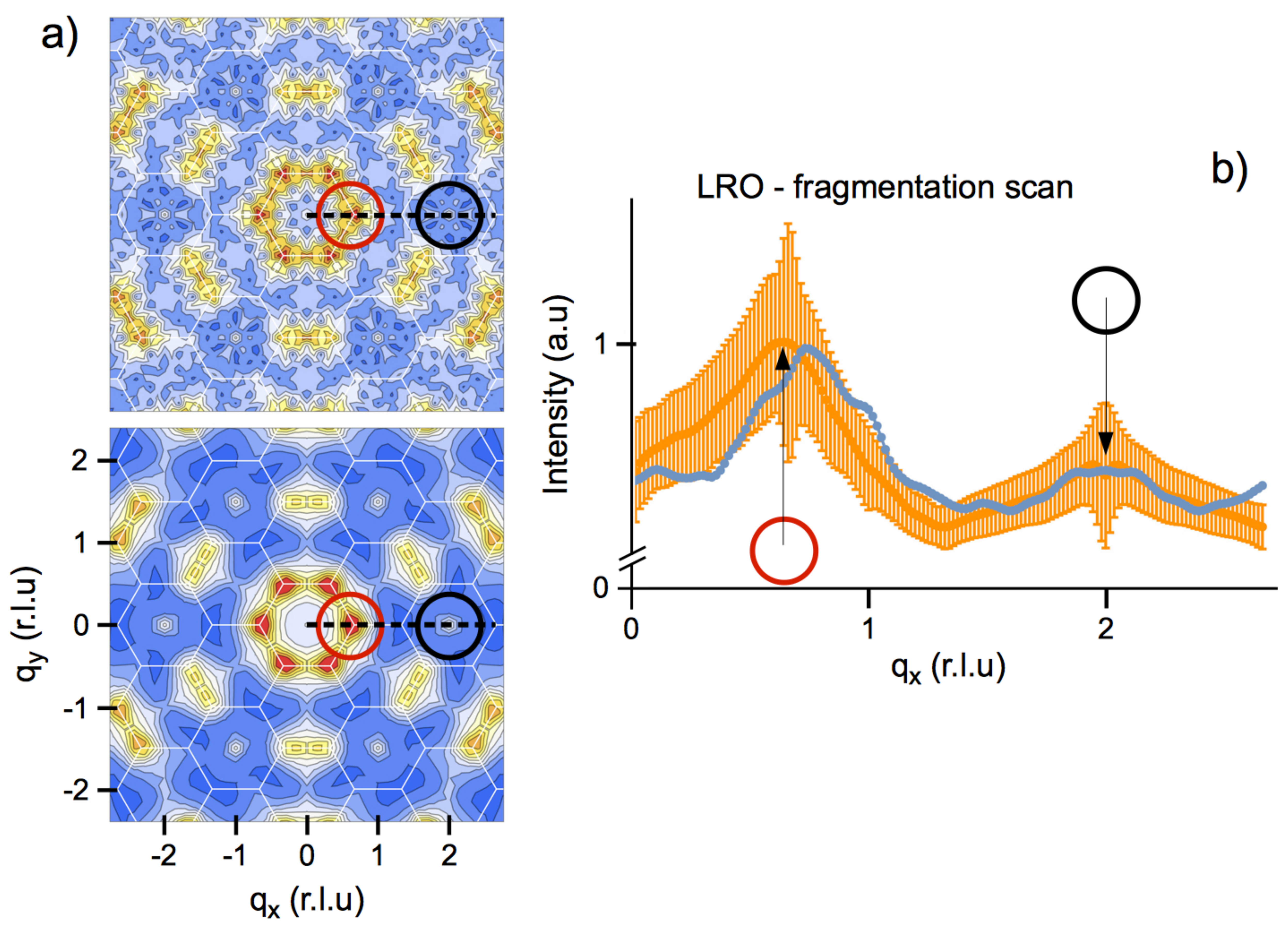}

\caption{{\bf Experimental and theoretical spin fragmentation in reciprocal space.} (a) Experimental (top) and theoretical (bottom) 2D maps of the magnetic structure factor corresponding to the experimental XMCD-PEEM image. 
The positions of the Bragg peaks corresponding to both the fragmented-spin phase and to the LRO are indicated by black and red circles, respectively. One particular scan in the reciprocal space is indicated by a dashed black line in the two maps. The scan starts at the origin of the reciprocal space and passes firstly through a LRO Bragg peak, and then through a fragmentation peak. (b) Comparison between the experimental (blue) and theoretical (orange) q-scans along the direction mentioned before. Standard deviations of the theoretical fluctuations (orange) are reported to quantify the likelihood of the dipolar spin ice model to describe the experimental observations. The dipolar spin ice model captures most features of spin-spin correlations and agrees semi-quantitatively on the amplitude, as well as on the positions, of the correlations. Intensity is given in arbitrary units but it must be noted that both experimental and theoretical curves have been scaled in a similar way, i.e. there is no free parameter but the effective temperature ($T/J_{nn}=0.051$) of the Monte Carlo simulations.
\label{fig:exp_vs_theo}}
 
\end{figure}

\begin{figure}
\centering
\includegraphics[width=0.7\linewidth]{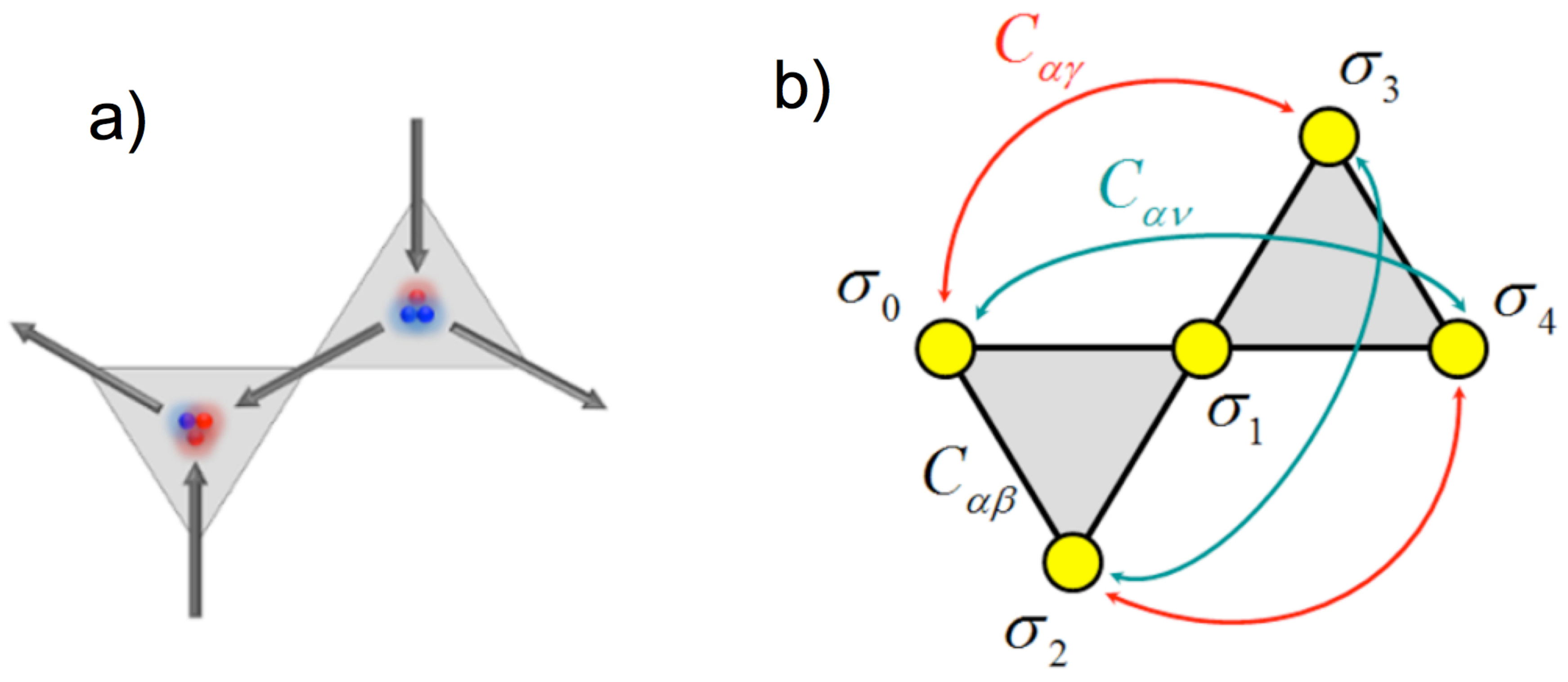}

\caption{{\bf Definition of the magnetic charges at the vertices and of the spin-spin correlations.} Every spin is the connection point between two vertex charges and contributes with a +1 magnetic charge in one triangle and -1 in the other. (a) A ($\sigma_0$,$\sigma_1$,$\sigma_2$,$\sigma_3$,$\sigma_4$) = ($-1,+1,+1,-1,+1$) configuration for the kagome spin ice. (b) In the general case, the vertex charges can be found by summing up the spin scalar values per triangle ($\sigma_i$). The constituting spins of a charge correlation pair form $\alpha\beta$, $\alpha\gamma$ and $\alpha\nu$ pairs.
\label{fig:Supp1}}

\end{figure}

\end{widetext}

\end{document}